\documentclass[
  aps,                  
  prd,                  
  reprint,              
  nofootinbib,          
  superscriptaddress,   
  showpacs,             
  showkeys,             
  preprintnumbers,      
  longbibliography      
]{revtex4-1}

\usepackage[utf8]{inputenc}
\usepackage{amsfonts,amsmath,amssymb}
\usepackage{bm,dsfont} 
\usepackage{url}
\usepackage{graphicx}
\usepackage{hyperref}
\usepackage{multirow}

\newcommand{\nwse}[3]{\ensuremath{#1^{#2}_{\phantom{#2} #3}}}
\newcommand{\swne}[3]{\ensuremath{#1_{#2}^{\phantom{#2} #3}}}

\newcommand{\bring}[1]{\mathring{\bar{#1}}}

\newcommand{\asctp}{Arnold Sommerfeld Center for Theoretical Physics, Ludwig-Maximilians-Universität München, Theresienstraße 37, 80333 Munich, Germany}
\newcommand{\udec}{Departamento de Fí­sica, Universidad de Concepción, Casilla 160-C, Concepción, Chile}
\newcommand{\unal}{Departamento de Fí­­sica, Universidad Nacional de Colombia, Bogotá, Colombia}
\newcommand{\unap}{Facultad de Ingeniería y Arquitectura, Universidad Arturo Prat, Iquique, Chile}
\newcommand{\ucn}{Departamento de Enseñanza de las Ciencias Básicas, Universidad Católica del Norte, Larrondo 1281, Coquimbo, Chile}

\begin{document}

\title{Nonminimal couplings, gravitational waves, and torsion in Horndeski's theory}

\affiliation{\unap}

\author{José Barrientos}
\email{josebarrientos@udec.cl}
\affiliation{\udec}
\affiliation{\ucn}

\author{Fabrizio Cordonier-Tello}
\email{f.cordonier@physik.uni-muenchen.de}
\affiliation{\asctp}

\author{Fernando Izaurieta}
\email{fizaurie@udec.cl}
\affiliation{\udec}

\author{Perla Medina}
\email{perlamedina@udec.cl}
\affiliation{\udec}

\author{Daniela Narbona}
\email{danielanarbona@udec.cl}
\affiliation{\udec}

\author{Eduardo Rodríguez}
\email{eduarodriguezsal@unal.edu.co}
\affiliation{\unal}
\affiliation{\asctp}

\author{Omar Valdivia}
\email{ovaldivi@unap.cl}
\affiliation{\unap}

\date{\today}

\begin{abstract}
  The Horndeski Lagrangian brings together all possible interactions
  between gravity and a scalar field that yield second-order field equations
  in four-dimensional spacetime.
  As originally proposed, it only addresses phenomenology without torsion,
  which is a non-Riemannian feature of geometry.
  Since torsion can potentially affect interesting phenomena such as
  gravitational waves and early Universe inflation, in this paper we allow
  torsion to exist and propagate within the Horndeski framework.
  To achieve this goal, we cast the Horndeski Lagrangian in Cartan's
  first-order formalism, and introduce wave operators designed to act
  covariantly on $p$-form fields that carry Lorentz indices.
  We find that nonminimal couplings and second-order derivatives of the
  scalar field in the Lagrangian are indeed generic sources of torsion.
  Metric perturbations couple to the background torsion and new torsional
  modes appear. These may be detected via gravitational waves but not
  through Yang--Mills gauge bosons.
\end{abstract}



\preprint{LMU-ASC 42/17}

\maketitle

%
%

\section{Introduction}


Recently, there has been a surge of interest
in Horndeski's theory~\cite{Horn74,Cha11,Nico08,Koba11,Bhattacharya:2016naa},
which is the most general four-dimensional scalar-tensor theory of gravity,
without torsion, that has second-order field equations.

Torsion, however, can have dramatic effects
in the very early universe~\cite{Pop10}, which is precisely one regime
where scalar fields are thought to be relevant.

In the Einstein--Cartan--Sciama--Kibble (ECSK) theory~\cite{Hehl76},
torsion is generated by fermions and affects only fermions.
Its effects are in general very weak,
since torsional terms are proportional to $\psi^{4}$
and hence only important when there is a large fermion density
(see section~8.4 of Ref.~\cite{SupergravityVanProeyen},
and Ref.~\cite{Kerlick1975}).
Torsional effects of this magnitude will likely go undetected in
any foreseeable particle physics experiment.
They may be detectable in cosmological scenarios~\cite{Puetzfeld:2004yg}
and in theories that go beyond ECSK in four dimensions
(see, e.g.,
Refs.~\cite{%
Sezgin1980,
Sezgin1981,
Chang:2000yw,
Lebedev:2002dp,
Jackiw:2003pm,
Obukhov:2006gea,
Cantcheff:2008qn,
Ertem:2009ur,
BaeklerHehl2011,
Hehl2013gauge,
Tol13,
Castillo-Felisola:2013jva,
Castillo-Felisola:2014iia,
Castillo-Felisola:2014xba%
}).

Standard model bosons, on the other hand,
do not generate and are not affected by torsion.%
\footnote{\label{ft:YM}This holds true when Yang--Mills bosons are described
mathematically by connections on principal bundles,
meaning, in particular, that the field strength reads
$F_{\mu\nu} = \partial_{\mu} A_{\nu} - \partial_{\nu} A_{\mu}
 + \frac{1}{2} \left[ A_{\mu}, A_{\nu} \right]$,
\emph{not}
$F_{\mu\nu} = \nabla_{\mu} A_{\nu} - \nabla_{\nu} A_{\mu}
+ \frac{1}{2} \left[ A_{\mu}, A_{\nu} \right]$,
which differs from the former when torsion is present.
Both points of view have been studied in the literature.
Some examples of the first can be found in
Refs.~\cite{Hehl76,Hammond2002,Shapiro:2001rz,Benn:1980ea}.
Some examples of the second point of view, coupling YM bosons and torsion,
can be found in Refs.~\cite{Novello:1976tx,Hojman1978,DeSabbata:1980gb}.
In this paper we only deal with the first approach.}

Most strikingly, torsion is a nonpropagating field.
In vacuum, the ECSK theory gives zero torsion,
so there can be no torsional modes for gravitational waves.

The kind of nonminimal couplings between gravity and a scalar field
that appear in Horndeski's theory can modify the conclusions drawn
from the ECSK theory.
Several authors
(see, e.g., Refs.~\cite{Jackiw:2003pm,Cantcheff:2008qn,Ertem:2009ur,Tol13})
have studied the consequences of including in the action
a term that is the product of a scalar field and the Euler four-form density,
\begin{equation}
  L = \phi \epsilon_{abcd} R^{ab} \wedge R^{cd},
  \label{eq:TZ}
\end{equation}
where $R^{ab}$ stands for the Lorentz curvature two-form.
Here they find a twofold surprise: contrary to expectations,
the term in eq.~(\ref{eq:TZ}) produces nontrivial dynamics
(in stark contrast with the uncoupled Euler density)
with torsion (which normally requires fermionic fields).
This term also appears naturally in several contexts
(see Refs.~\cite{Cha90,Salgado:2013pva,Salgado:2014jka,Salgado:2014iha,Catalan:2015bec}).


In this paper, we study Horndeski's theory \emph{with} torsion, i.e.,
a theory whose relation to the original Horndeski theory is the same
as that of the ECSK theory to general relativity.

Our main result can be stated as follows:
Nonminimal couplings between gravity and a scalar field generically produce torsion.
This nontrivial torsion can be thought of as an effective dark matter,
which may in principle be detected via gravitational waves
but not through Yang--Mills (YM) gauge bosons.

Since torsion is a non-Riemannian feature of geometry, we find it convenient
to work with Cartan's differential geometry formalism.
In section~\ref{sec:1OFHT} we setup some useful definitions to deal with
the Horndeski Lagrangian in the first-order formalism.
We write down the Horndeski Lagrangian in its most general form and deduce its
field equations from the variations with respect to the independent fields:
the vierbein $e^{a}$, the spin connection $\omega^{ab}$, and the scalar field $\phi$.
An interesting analysis regarding the phenomenology between
some nonminimal couplings and torsion components can be found in Ref.~\cite{Sur:2016bwu}.
The main difference between the section~\ref{sec:1OFHT} of current work
and Ref.~\cite{Sur:2016bwu} is that in the current article we consider
the full Horndeski Lagrangian, and in Ref.~\cite{Sur:2016bwu} they explore
the idea of torsion as dark energy in cosmological models. 

Before plunging into an in-depth analysis of first-order perturbation theory,
in section~\ref{sec:waveops} we pause for a moment in order to study
which torsion-aware wave operators are the most appropriate to act on
our fields, which are in general differential forms that carry Lorentz indices.
In this mostly mathematical section we provide a generalized version of the
Weitzenböck identity that relates torsion-aware versions of the Laplace--Beltrami
and the Laplace--de~Rham operators, which may have an interest of its own.

In section~\ref{sec:GWTM} we establish the first-order perturbation theory
for a theory of gravity, in its first-order formalism guise, and a scalar field,
and apply it to the most interesting bits of the Horndeski Lagrangian---namely,
those that can lead to gravitational waves.
This perturbation theory is nontrivial because, to the best of our knowledge,
up until now it has been unclear how to separate the metric from the torsional
degrees of freedom in the first-order perturbation of the spin connection
on backgrounds with curvature and torsion. For flat backgrounds, linearized gravity in first-order formalism can be found in Ref.~\cite{Baykal:2016nek}. 
The separation that we achieve in eqs.~(\ref{eq:Udef})--(\ref{eq:Vdef})
is novel and, while it serves as a useful step in establishing our main
result---nonminimal couplings lead to nontrivial torsion that ``hitches a ride''
on gravitational waves---, may also find applications elsewhere.

For the sake of simplicity, in this paper we restrict ourselves to generalizing
Horndeski's theory to allow for nonzero torsion, without ever going beyond
second-order field equations. There exist, however, healthy theories that manage
to evade the Ostrogradski instability while including higher-order derivatives
(see, e.g., Refs.~\cite{Gleyzes:2014dya,Langlois:2015cwa,BenAchour:2016fzp}).
The addition of torsion to this kind of theories remains an open problem.


\section{First order formalism for Horndeski's theory}
\label{sec:1OFHT}

In this section we analyze the general behavior of the Horndeski Lagrangian
without imposing the torsionless condition.
The result found in Ref.~\cite{Tol13} for the Gauss--Bonnett term coupled to a scalar field
[cf.~eq.~(\ref{eq:TZ})] proves to be a general feature,
and nonminimal couplings with an scalar field are shown to be sources of torsion.


\subsection{Preliminaries}

Let's begin with some definitions.

We shall take spacetime to be a four-dimensional smooth manifold
$M$ with signature $\left( -+++ \right)$.
Greek indices $\mu,\nu,\ldots=0,1,2,3$ are used for tensor components
in the coordinate basis, while lower-case Latin indices $a,b,\ldots=0,1,2,3$
are used for the Lorentz (orthonormal) basis.
The components of the change-of-basis matrix, $\nwse{e}{a}{\mu}$,
help us define the one-form vierbein as
$e^{a} = \nwse{e}{a}{\mu} \mathrm{d}x^{\mu}$.
The spacetime metric $g_{\mu\nu}$ can be written as
\begin{equation}
  ds^{2} =
  g_{\mu \nu} \mathrm{d}x^{\mu} \otimes \mathrm{d}x^{\nu} =
  \eta_{ab} e^{a} \otimes e^{b},
\end{equation}
whence $g_{\mu\nu} = \eta_{ab} \nwse{e}{a}{\mu} \nwse{e}{b}{\nu}$.
The space of all $p$-forms defined on $M$
is denoted as $\Omega^{p} \left( M \right)$.

It proves useful to define an operator $\Sigma_{a_{1}\cdots a_{q}}$
that maps $p$-forms into $\left( p-q \right)$-forms,
\begin{equation}
  \Sigma_{a_{1}\cdots a_{q}} :
  \Omega^{p} \left( M \right) \rightarrow
  \Omega^{p-q} \left( M \right) ,
\end{equation}
and is defined by its action on a $p$-form $\alpha$ as
\begin{equation}
  \Sigma^{a_{1} \cdots a_{q}} \alpha =
  - \left( -1 \right)^{p \left( p - q \right)} \ast \left(
    e^{a_{1}} \wedge \cdots \wedge e^{a_{q}} \wedge
    \ast \alpha
  \right).
\end{equation}
Here, $\ast$ stands for the Hodge dual,
which maps $p$-forms into $\left( 4-p \right)$-forms,
$\ast: \Omega^{p} \left( M \right) \rightarrow
\Omega^{4-p} \left( M \right)$.

When $q=1$ we find
\begin{equation}
  \Sigma^{a} \alpha =
  - \ast \left( e^{a} \wedge \ast \alpha \right).
\end{equation}
This case is particularly interesting, since $\Sigma_{a}$
behaves as an exterior derivative:
(i)~it satisfies Leibniz's rule,
\begin{equation}
  \Sigma_{a} \left(
    \alpha \wedge \beta
  \right) =
  \Sigma_{a} \alpha \wedge \beta +
  \left( -1 \right)^{p}
  \alpha \wedge \Sigma_{a} \beta,    
\end{equation}
and (ii)~is nilpotent,
\begin{equation}
  \Sigma_{a} \Sigma^{a} = 0.    
\end{equation}
A key difference between $\Sigma_{a}$ and $\mathrm{d}$ is that,
while $\mathrm{d}$ \emph{increases} the degree of a differential form by one,
$\Sigma_{a}$ \emph{decreases} it by the same amount.

In order to write the Horndeski Lagrangian in first-order formalism
(and not impose the torsionless condition from the beginning%
\footnote{The \textit{torsionless} Horndeski theory has already been studied
in the language of differential forms; see Ref.~\cite{Ezquiaga:2016nqo}.}),
we will describe the geometry by means of the vierbein one-form $e^{a}$,
the one-form spin connection $\omega^{ab}$, and the scalar field $\phi$.
The spin connection and the vierbein represent independent degrees of freedom,
and torsion and Lorentz curvature two-forms are given by
\begin{align}
  T^{a} & = \mathrm{D} e^{a} =
  \mathrm{d} e^{a} + \nwse{\omega}{a}{b} \wedge e^{b}, \\
  R^{ab} & = \mathrm{d} \omega^{ab} +
  \nwse{\omega}{a}{c} \wedge \omega^{cb}.
\end{align}

A small circle above a quantity will be used to denote the
``torsionless version'' of that quantity.
For instance, the spin connection can always be split as
\begin{equation}
  \omega^{ab} = \mathring{\omega}^{ab} + \kappa^{ab},    
\end{equation}
where $\mathring{\omega}^{ab}$ stands for the usual torsion-free
one-form spin connection derived from the vierbein,
and $\kappa^{ab}$ is the one-form \emph{contorsion}.
In the same way, the Lorentz curvature two-form can be expressed as
\begin{equation}
  R^{ab} = \mathring{R}^{ab} + \mathrm{\mathring{D}} \kappa^{ab} +
  \nwse{\kappa}{a}{c} \wedge \kappa^{cb},
  \label{Eq_Lorentz_Curvature}
\end{equation}
where $\mathring{R}^{ab}$ is the torsion-independent two-form Riemann curvature,
$\mathring{R}^{ab} = \mathrm{d} \mathring{\omega}^{ab} +
 \nwse{\mathring{\omega}}{a}{c} \wedge \mathring{\omega}^{cb}$,
and $\mathring{\mathrm{D}}$ stands for the exterior covariant derivative
with respect to the torsion-free connection $\mathring{\omega}^{ab}$.

In order to deal with the scalar field $\phi$ and its derivatives
in this first-order formalism context, it proves useful to define the zero-form
\begin{equation}
  Z^{a} = \Sigma^{a} \mathrm{d} \phi,    
\end{equation}
and the one-forms
\begin{align}
  \pi^{a} & = \mathrm{D} Z^{a},\\
  \theta^{a} & = Z^{a} \mathrm{d} \phi.
\end{align}
Intuitively, one can think of $Z^{a}$ as the derivative of $\phi$
in the direction specified by the $a$-index, while
$\pi^{a}$ and $\theta^{a}$ represent $\partial^{2}\phi$ and $\left(\partial\phi\right)^{2}$, respectively.


\subsection{The Horndeski Lagrangian}

Using the $\Sigma_{a}$ operator and its properties,
it is straightforward to work with the Horndeski Lagrangian
in the first-order formalism.
For instance, the ``Fab Four'' Lagrangians from Ref.~\cite{Cha11}
can be rewritten as the four-forms
\begin{align}
  L_{\text{J}} & =
  \frac{1}{2} V_{\text{J}} \left( \phi \right)
  \epsilon_{abcd} R^{ab} \wedge e^{c} \wedge \theta^{d},
  \\
  L_{\text{P}} & =
  \frac{1}{2} V_{\text{P}} \left( \phi \right)
  \epsilon_{abcd} R^{ab} \wedge \theta^{c} \wedge \pi^{d},
  \\
  L_{\text{G}} & =
  \frac{1}{2} V_{\text{G}} \left( \phi \right)
  \epsilon_{abcd} R^{ab} \wedge e^{c} \wedge e^{d},
  \\
  L_{\text{R}} & =
  \frac{1}{2} V_{\text{R}} \left( \phi \right)
  \epsilon_{abcd} R^{ab} \wedge R^{cd}.
\end{align}
The \textit{Ringo} and \textit{George} cases are straightforward
to translate from tensor language into differential forms,
while the \textit{John} and \textit{Paul} cases prove more interesting.
In particular, for \textit{Paul} it is much more comfortable to work with $\theta^{a}$ and $\pi^{a}$ instead of the Riemann double-dual.

The same is true for the full Horndeski Lagrangian.
In terms of the variables we have defined,
the Horndeski Lagrangian four-form reads
\begin{align}
  L_{\text{H}} \left( \phi, e, \omega \right) & =
  \epsilon_{abcd} \left(
    2 \kappa_{1} R^{ab} \wedge e^{c} \wedge \pi^{d} +
    \right. \nonumber \\ & +
    \frac{2}{3} \frac{\partial \kappa_{1}}{\partial X} \pi^{a}
    \wedge \pi^{b} \wedge\pi^{c} \wedge e^{d} +
    \nonumber \\ & +
    2 \kappa_{3} R^{ab} \wedge e^{c} \wedge \theta^{d} +
    \nonumber \\ & +
    2 \frac{\partial \kappa_{3}}{\partial X} \theta^{a}
    \wedge \pi^{b} \wedge \pi^{c} \wedge e^{d} +
    \nonumber \\ & +
    \left( F + 2W \right) R^{ab} \wedge e^{c} \wedge e^{d} +
    \nonumber \\ & +
    \frac{\partial F}{\partial X} \pi^{a}
    \wedge \pi^{b} \wedge e^{c} \wedge e^{d} +
    \nonumber \\ & +
    \kappa_{8} \theta^{a} \wedge \pi^{b} \wedge e^{c} \wedge e^{d} +
    \nonumber \\ & -
    \left[
      \frac{\partial \left( F + 2W \right)}{\partial \phi} - X \kappa_{8}
    \right]
    \pi^{a} \wedge e^{b} \wedge e^{c} \wedge e^{d} +
    \nonumber \\ & +
    \left. \kappa_{9} \frac{1}{4!} e^{a} \wedge e^{b} \wedge e^{c} \wedge e^{d}
  \right),
  \label{Ec_Horndeski_Formas}
\end{align}
where the arbitrary functions ($i = 1, 3, 8, 9$)
\begin{align}
 \kappa_{i} & = \kappa_{i} \left( \phi, X \right), \\
 F & = F \left( \phi, X \right), \\
 W & = W \left( \phi \right),
\end{align}
must satisfy the constraint
\begin{equation}
 \mathcal{C} \left( \phi, X \right) =
 \frac{\partial F}{\partial X} - 2 \left(
   \kappa_{3} + 2X \frac{\partial \kappa_{3}}{\partial X} -
   \frac{\partial \kappa_{1}}{\partial \phi}
 \right)
 = 0,
 \label{Eq_Constraint}
\end{equation}
with
\begin{equation}
  X = - \frac{1}{2} Z_{a} Z^{a}.    
\end{equation}

It is interesting to notice that the Hodge $\ast$-operator appears in
the Horndeski Lagrangian exclusively through the $\Sigma^{a}$ operator.
This operator allows us to cast the full Horndeski Lagrangian
in an effective Lovelock-like mold~\cite{Lov71,Zum86},
with the Lorentz one-forms $\pi^{a}$ and $\theta^{a}$
playing a role similar to that of the vierbein, $e^{a}$.

Eq.~(\ref{Ec_Horndeski_Formas}) gives the full Horndeski Lagrangian
in Cartan's first-order formalism. The Horndeski Theorem~\cite{Horn74}
states that, when torsion vanishes, this is the most general scalar-tensor
Lagrangian that gives rise to second-order equations for the metric.
When torsion is allowed to exist, however, Horndeski's theorem is no longer valid.
Indeed, it is quite easy to come up with new terms, explicitly involving
torsion, that don't spoil the second-order nature of the field equations.
For the sake of simplicity, in this article we will concern ourselves
solely with the Horndeski Lagrangian as shown in eq.~(\ref{Ec_Horndeski_Formas}).
The generalization of the Horndeski theorem
for the case of nonvanishing torsion, i.e.,
the answer to the question
``What is the most general Lagrangian that leads to second-order
field equations for the metric on a spacetime with torsion?''
remains an open problem and will be considered elsewhere.


\subsection{Field Equations}

In order to derive the field equations in the first-order paradigm,
we treat $\omega^{ab}$, $e^{a}$, and $\phi$ as independent degrees of freedom.%
\footnote{Note that $Z^{a}$ depends on $e^{a}$ and the derivatives of $\phi$
through the $\Sigma^{a}$ operator. This dependence must be taken into account
when performing the variations with respect to $e^{a}$ and $\phi$.}

Explicitly performing the variation with respect to
the spin connection yields the three-form equation
$\mathcal{E}_{ab} = 0$, where
\begin{align}
  \mathcal{E}_{ab} & = -
  \epsilon_{abcd} T^{c} \wedge \left[
    \kappa_{1} \pi^{d} + \kappa_{3} \theta^{d} +
    \left( F + 2W \right) e^{d}
  \right] +
  \nonumber \\ & +
  \epsilon_{abcd} e^{c} \wedge \left[
    \mathrm{d} \kappa_{1} \wedge \pi^{d} + \kappa_{1} R^{d}{}_{e} Z^{e} +
    \mathrm{d} \kappa_{3} \wedge \theta^{d} +
  \right.
  \nonumber \\ & \left.
    -\kappa_{3} \mathrm{d} \phi \wedge \pi^{d} +
    \frac{1}{2} \mathrm{d} \left( F + 2 W \right) \wedge e^{d}
  \right] +
  \nonumber \\ & -
  \frac{1}{2} \left( Z_{a} \epsilon_{bcde} - Z_{b} \epsilon_{acde} \right)
  \left[
    \kappa_{1} R^{cd} +
  \right.
  \nonumber \\ & +
  \pi^{c} \wedge \left(
    \frac{\partial \kappa_{1}}{\partial X} \pi^{d} +
    2\frac{\partial\kappa_{3}}{\partial X} \theta^{d} +
    \frac{\partial F}{\partial X} e^{d}
  \right) +
  \nonumber \\ & \left. +
    \frac{1}{2} \left(
      \kappa_{8} \theta^{c} -
      \left[
        \frac{\partial}{\partial \phi} \left( F + 2 W \right) - X \kappa_{8} 
      \right]
      e^{c}
    \right)
    \wedge e^{d}
  \right]
  \wedge e^{e}
  \label{Eq_Motion_wab}
\end{align}

The field equations obtained from variation with respect to
the vierbein and the scalar field, on the other hand, read
\begin{align}
  \mathcal{E}_{a} & = E_{a} +
  \Sigma^{b} \left(
    \mathcal{S}_{b} + \mathcal{T}_{b} + \mathcal{U}_{b}
  \right)
  Z_{a} = 0,
  \label{Eq_Motion_ea} \\
  \mathcal{E} & =
  E + \mathcal{Z} -
  \mathrm{d} \Sigma^{b} \left(
    \mathcal{S}_{b} + \mathcal{T}_{b} + \mathcal{U}_{b}
  \right)
  = 0,
  \label{Eq_Motion_phi}
\end{align}
where
\begin{align}
  E_{d} & = \epsilon_{abcd} \left(
    2 \kappa_{1} R^{ab} \wedge \pi^{c} +
    \frac{2}{3} \frac{\partial \kappa_{1}}{\partial X}
    \pi^{a} \wedge \pi^{b} \wedge \pi^{c} +
  \right. \nonumber \\ & + 
  2 \kappa_{3} R^{ab} \wedge \theta^{c} +
  2 \frac{\partial \kappa_{3}}{\partial X}
  \theta^{a} \wedge \pi^{b} \wedge \pi^{c} +
  \nonumber \\ & +
  2 \left( F + 2 W \right) R^{ab} \wedge e^{c} +
  2 \frac{\partial F}{\partial X} \pi^{a} \wedge \pi^{b} \wedge e^{c} +
  \nonumber \\ & +
  2 \kappa_{8} \theta^{a} \wedge \pi^{b} \wedge e^{c} +
  \frac{1}{3!} \kappa_{9} e^{a} \wedge e^{b} \wedge e^{c} +
  \nonumber \\ & \left. -
    3 \left[
      \frac{\partial}{\partial\phi} \left( F + 2 W \right) - X \kappa_{8}
    \right]
    \pi^{a} \wedge e^{b} \wedge e^{c}
  \right),
\end{align}
\begin{align}
  E & = \epsilon_{abcd} \left[
    2 \left( \frac{\partial \kappa_{1}}{\partial \phi} - \kappa_{3} \right)
    R^{ab} \wedge e^{c} \wedge \pi^{d}
  \right. +
  \nonumber \\ & +
  2 \left(
    \frac{1}{3} \frac{\partial^{2} \kappa_{1}}{\partial \phi \partial X} -
    \frac{\partial \kappa_{3}}{\partial X}
  \right)
  \pi^{a} \wedge \pi^{b} \wedge \pi^{c} \wedge e^{d} +
  \nonumber \\ & +
  2 \frac{\partial \kappa_{3}}{\partial \phi}
  R^{ab} \wedge e^{c} \wedge \theta^{d} +
  \nonumber \\ & +
  2 \frac{\partial^{2} \kappa_{3}}{\partial \phi \partial X}
  \theta^{a} \wedge \pi^{b} \wedge \pi^{c} \wedge e^{d} +
  \nonumber \\ & + \left(
    \frac{\partial F}{\partial \phi} + 2 \frac{\partial W}{\partial \phi}
  \right)
  R^{ab} \wedge e^{c} \wedge e^{d} +
  \nonumber \\ & + \left(
    \frac{\partial^{2} F}{\partial \phi \partial X} - \kappa_{8}
  \right)
  \pi^{a} \wedge \pi^{b} \wedge e^{c} \wedge e^{d} +
  \nonumber \\ & +
  \frac{\partial \kappa_{8}}{\partial \phi}
  \theta^{a} \wedge \pi^{b} \wedge e^{c} \wedge e^{d} +
  \nonumber \\ & - \left[
    \frac{\partial^{2} \left( F+2W \right)}{\partial \phi^{2}} -
    X \frac{\partial \kappa_{8}}{\partial \phi}
  \right]
  \pi^{a} \wedge e^{b} \wedge e^{c} \wedge e^{d} +
  \nonumber \\ & \left. +
    \frac{1}{4!} \frac{\partial \kappa_{9}}{\partial \phi}
    e^{a} \wedge e^{b} \wedge e^{c} \wedge e^{d}
  \right],
\end{align}
\begin{align}
  \mathcal{Z} & = \left[
    2 \mathrm{d} \kappa_{3} \wedge R^{ab} +
    2 \mathrm{d} \frac{\partial \kappa_{3}}{\partial X}
    \wedge \pi^{a} \wedge \pi^{b} +
  \right.
  \nonumber \\ & +
  \mathrm{d} \kappa_{8} \wedge \pi^{a} \wedge e^{b} +
  \nonumber \\ & \left. +
    \mathrm{D} \pi^{a} \wedge \left(
      4 \frac{\partial\kappa_{3}}{\partial X} \pi^{b} + \kappa_{8} e^{b}
    \right)
  \right]
  \wedge e^{c} Z^{d} +
  \nonumber \\ & +
  2 \epsilon_{abcd} \left[
    \kappa_{3} R^{ab} +
    \frac{\partial \kappa_{3}}{\partial X} \pi^{a} \wedge \pi^{b} +
  \right.
  \nonumber \\ & \left. +
    \kappa_{8} \pi^{a} \wedge e^{b}
  \right]
  \wedge T^{c}Z^{d},
\end{align}
and the common 4-form variables
$\mathcal{S}_{a}$, $\mathcal{T}_{a}$ and $\mathcal{U}_{a}$
are given by
\begin{align}
  \mathcal{S}_{d} & = 2 \epsilon_{abcd} \left[
    \mathrm{D} \pi^{a} \wedge e^{b} \wedge \left(
      2 \frac{\partial \kappa_{1}}{\partial X} \pi^{c} +
      2 \frac{\partial \kappa_{3}}{\partial X} \theta^{c} +
      \frac{\partial F}{\partial X} e^{c}
    \right)
  \right.
  + \nonumber \\ & +
  \pi^{a} \wedge e^{b} \wedge \mathrm{d} X \wedge \left(
    \frac{\partial^{2} \kappa_{1}}{\partial X^{2}} \pi^{c} +
    2 \frac{\partial^{2} \kappa_{3}}{\partial X^{2}} \theta^{c} +
    \frac{\partial^{2} F}{\partial X^{2}} e^{c}
  \right)
  + \nonumber \\ & + \left.
    \frac{1}{2} e^{a} \wedge e^{b} \wedge \mathrm{d} X \wedge \left(
      \theta^{c} \frac{\partial \kappa_{8}}{\partial X} -
      e^{c} \frac{\partial}{\partial X} \left\{
        \frac{\partial F}{\partial \phi} - X \kappa_{8}
      \right\}
    \right)
  \right],
  \label{Eq_Sa}
\end{align}
\begin{align}
  \mathcal{T}_{d} & = 2 \epsilon_{abcd} \left[
    \kappa_{1} R^{ab} +
    \frac{\partial \kappa_{1}}{\partial X} \pi^{a} \wedge \pi^{b} +
    2 \frac{\partial \kappa_{3}}{\partial X} \pi^{a} \wedge \theta^{b} +
  \right.
  \nonumber \\ & +
  2 \frac{\partial F}{\partial X} \pi^{a} \wedge e^{b} +
  \frac{1}{2} \kappa_{8} e^{a} \wedge \theta^{b} +
  \nonumber \\ & \left. -
    \frac{3}{2} \left(
      \frac{\partial}{\partial\phi} \left( F + 2 W \right) - X \kappa_{8}
    \right)
    e^{a} \wedge e^{b}
  \right]
  \wedge T^{c},
\end{align}
\begin{align}
  \mathcal{U}_{e} & = \epsilon_{abcd} \left[
    -R^{ab} \wedge e^{c} \wedge \left(
      \nwse{C}{d}{e} +
      2 \frac{\partial \kappa_{1}}{\partial X} \delta_{ef}^{gd} Z_{g} \pi^{f}
    \right)
    +
  \right.
  \nonumber \\ & -
  \pi^{a} \wedge \pi^{b} \wedge e^{c} \wedge \left(
    \nwse{\bar{C}}{d}{e} +
    \frac{2}{3} \frac{\partial^{2} \kappa_{1}}{\partial X^{2}} \pi^{d} Z_{e}
  \right)
  + \nonumber \\ & \left. +
    \pi^{a} \wedge e^{b} \wedge e^{c} \wedge \nwse{M}{d}{e} +
    e^{a} \wedge e^{b} \wedge e^{c} \wedge \nwse{K}{d}{e}
  \right].
  \label{eq:defU_a}
\end{align}
In eq.~(\ref{eq:defU_a}), 
$\nwse{C}{a}{b}$,
$\nwse{\bar{C}}{a}{b}$,
$\nwse{K}{a}{b}$, and 
$\nwse{M}{a}{b}$
are one-forms defined as
\begin{align}
  \nwse{C}{a}{b} & = 2 \mathrm{d} \phi \left[
    \frac{\partial \kappa_{3}}{\partial X} Z^{a} Z_{b} -
    \left( \kappa_{3} - \frac{\partial \kappa_{1}}{\partial \phi} \right)  \delta_{b}^{a}
  \right]
  + \nonumber \\ & +
  e^{a} Z_{b} \frac{\partial F}{\partial X},
  \\
  \nwse{\bar{C}}{a}{b} & = 2 \mathrm{d} \phi \left[
    \frac{\partial^{2} \kappa_{3}}{\partial X^{2}} Z^{a} Z_{b} -
    \left(
      3 \frac{\partial \kappa_{3}}{\partial X} -
      \frac{\partial^{2} \kappa_{1}}{\partial \phi \partial X}
    \right)
    \delta_{b}^{a}
  \right]
  + \nonumber \\ & +
  e^{a} Z_{b} \frac{\partial^{2} F}{\partial X^{2}},
  \\
  \nwse{K}{a}{b} & = \left[
    \frac{\partial^{2}}{\partial \phi^{2}} \left( F + 2 W \right) -
    X \frac{\partial \kappa_{8}}{\partial \phi}
  \right]
  \mathrm{d} \phi \delta_{b}^{a} +
  \nonumber \\ & -
  \frac{1}{4!} e^{a} Z_{b} \frac{\partial \kappa_{9}}{\partial X},
  \\
  \nwse{M}{a}{b} & = \left(
    2 \left[
      \kappa_{8} - \frac{\partial^{2} F}{\partial \phi\partial X}
    \right]
    \delta_{b}^{a} -
    \frac{\partial \kappa_{8}}{\partial X} Z^{a} Z_{b}
  \right)
  \mathrm{d} \phi +
  \nonumber \\ & +
  e^{a} Z_{b} \frac{\partial}{\partial X} \left[
    \frac{\partial F}{\partial \phi} - X \kappa_{8}
  \right].
  \label{eq:Mdef}
\end{align}

The one-forms $\nwse{C}{a}{b}$ and $\nwse{\bar{C}}{a}{b}$
satisfy the properties
\begin{align}
 \Sigma^{b} \nwse{C}{a}{b} & = Z^{a} \mathcal{C}, \\
 \Sigma^{b} \nwse{\bar{C}}{a}{b} & =
 Z^{a} \frac{\partial \mathcal{C}}{\partial X},
\end{align}
where $\mathcal{C} \left(\phi, X \right) = 0$
is the Horndeski constraint~(\ref{Eq_Constraint}).

Here it is important to observe that in the terms
$\mathcal{Z}$ and $\mathcal{T}_{a}$
torsion appears explicitly as a result of nonminimal couplings.
Torsional degrees of freedom are also contained inside the Lorentz curvature
through the contorsion one-form, as shown in eq.~(\ref{Eq_Lorentz_Curvature}).

As the quickest glance at eqs.~(\ref{Eq_Motion_wab})--(\ref{eq:Mdef})
will show, the full Horndeski theory is extremely complicated.
Actually, it may be more accurate to think of it as a family of theories,
each one defined by a choice of the arbitrary functions $\kappa_{i}$,
$F$, and $W$. There are, however, several general observations to be made.

First, using the properties of the $\Sigma^{a}$ operator it is straightforward to derive the field equations obtained from the independent variations
of $e^{a}$ and $\omega^{ab}$, without imposing the torsionless condition.
Trying to achieve the same feat in the standard Palatini tensor formalism
would have been impractical, to say the least.

Second, the field equations directly show that torsion arises, in general,
from every nonminimal coupling with the scalar field,
and from the terms depending on
$\pi^{a} = \mathrm{D} \Sigma^{a} \mathrm{d} \phi$.
For instance, see the term $\mathcal{T}_{a}$ in eqs.~(\ref{Eq_Motion_ea}) and~(\ref{Eq_Motion_phi}), and the first term of eq.~(\ref{Eq_Motion_wab}).

In order to recover the standard torsionless dynamics,
one cannot simply impose $T^{a}=0$ on the equations of motion.
This happens because, in the general setting,
the dynamics of $\phi$ and the torsion become fully intertwined,
generically leading to $T^{a} \sim \partial \phi$.
Therefore, imposing $T^{a}=0$ in these cases will lead to
$\phi = \text{const.}$, freezing the dynamics of the scalar field.
The important point here is that the standard torsionless case
corresponds to a \emph{constraint} on the more general Cartan geometry framework.
This problem seems to have been known for a long time;
see, e.g., Ref.~\cite{Iglesias:2007nv},
or section~1.7.1 of Ref.~\cite{CastellaniSUGRA}.
The solution for it can be be written in a very practical way
in terms of the $\Sigma^{a}$ operator.
First, we have to include the torsionless condition via a two-form
Lagrange multiplier with a Lorentz index, $\Lambda_{a}$,
\begin{equation}
  L_{\text{H}} \rightarrow \bar{L}_{\text{H}} =
  L_{\text{H}} + \Lambda_{a} \wedge T^{a},    
\end{equation}
whence we get the new equations of motion
\begin{align}
  \bar{\mathcal{E}}_{a} & = \mathcal{E}_{a} - \mathrm{D} \Lambda_{a} = 0, \\
  \bar{\mathcal{E}} & = \mathcal{E} = 0, \\
  \bar{\mathcal{E}}^{ab} & = \mathcal{E}^{ab} - \frac{1}{2}
  \left( \Lambda^{a} \wedge e^{b} - \Lambda^{b} \wedge e^{a} \right) = 0, \\
  T^{a} & =0.
\end{align}
Using the $\Sigma^{a}$ operator, it is possible to solve
$\bar{\mathcal{E}}^{ab} = 0$ for $\Lambda_{a}$. We find
\begin{equation}
  \Lambda^{a} = 2 \Sigma_{b} \mathcal{E}^{ab} +
  \frac{1}{2} e^{a} \wedge \Sigma_{bc} \mathcal{E}^{bc}.
\end{equation}
Therefore, the standard field equations for the torsionless Horndeski theory
are recovered in this setting as
\begin{align}
  \mathcal{E}^{a} - 2 \mathrm{D} \Sigma_{b} \mathcal{E}^{ab} +
  \frac{1}{2} e^{a} \wedge \mathrm{d} \Sigma_{bc}
  \left. \mathcal{E}^{bc} \right\vert_{T^{a} = 0} & = 0, \\
  \left. \mathcal{E} \right\vert_{T^{a} = 0} & = 0.
\end{align}

This behavior is in stark contrast with the standard Einstein--Cartan case with minimally coupled fields. In this case, $T^{a}=0$ is an equation of motion in vacuum, and therefore it is unnecessary to use a Lagrange multiplier.
In fact, in this case only fermionic fields can be a source of non-propagating torsion (see, e.g., section~8.4 of Ref.~\cite{SupergravityVanProeyen}).

Known results for the torsionless Horndeski theory encompass
from cosmological models to black hole solutions.
In the torsional Horndeski setting here presented,
such solutions might persist~\cite{Jackiw:2003pm,Cantcheff:2008qn}
if one uplifts them in a consistent manner.
This means, in particular, that the Riemann curvature and the scalar field
should be the same as in the torsionless theory. Torsion, however, must be present
if the scalar field is nontrivial, and this will generically imply that the
Lorentz curvature shall differ from the Riemann curvature~\cite{Alv15,Cor15}.
Because of the complexity of Horndeski's theory, it seems unlikely
that one can make a general statement as to whether this uplifting
can always be done, or which conditions must be fulfilled for it to succeed.


\section{Wave operators, Torsion, and the Weitzenböck identity}
\label{sec:waveops}

Our goal in this section is to define a wave operator that can act on
differential forms that carry Lorentz indices, such as the vierbein, $e^{a}$.
We need this operator because our treatment of gravitational waves relies
on perturbations of the vierbein and the spin connection, $\omega^{ab}$,
which are the natural independent degrees of freedom for a spacetime with torsion.

\begin{table}[thpb]
  \caption{\label{tab:dbag}Many different derivatives are defined in this section. This table collects all definitions and some of their most important properties.}
  \begin{ruledtabular}
  \begin{tabular}{cccc}
    \multirow{2}{*}{Symbol} & \multirow{2}{*}{Definition} & Change in & Key\\
    & & form degree & property \\
    \hline
    $\mathrm{d}$ & $\mathrm{d}x^{\mu} \partial_{\mu}$ & $+1$ & $d^{2} = 0$ \\
    $\mathrm{D}$ & $\mathrm{d} + \omega$ & $+1$ & \\
    $\mathring{\mathrm{D}}$ & $\mathrm{d} + \mathring{\omega}$ & $+1$ & \\
    $\mathrm{d}^{\dagger}$ & $*\mathrm{d}*$ & $-1$ & \\
    $\mathrm{D}^{\dagger}$ & $*\mathrm{D}*$ & $-1$ & \\
    $\mathring{\mathrm{D}}^{\dagger}$ & $*\mathring{\mathrm{D}}*$ & $-1$ & \\
    $\Sigma^{a}$ & $-*\left( e^{a}\wedge *\right.$ & $-1$ & $\Sigma^{a} \Sigma_{a} = 0$ \\
    $\mathrm{D}^{\ddag}$ & $-\Sigma^{a} \mathrm{D} \Sigma_{a}$ & $-1$ & \\
    $\mathring{\mathrm{D}}^{\ddag}$ & $-\Sigma^{a} \mathring{\mathrm{D}} \Sigma_{a}$ & $-1$ & $\mathring{\mathrm{D}}^{\ddag} = \mathring{\mathrm{D}}^{\dagger}$ \\
    $\mathcal{D}_{a}$ & $\Sigma_{a} \mathrm{D} + \mathrm{D} \Sigma_{a}$ & 0 & \\
    $\mathring{\mathcal{D}}_{a}$ & $\Sigma_{a} \mathring{\mathrm{D}} + \mathring{\mathrm{D}} \Sigma_{a}$ & 0 & $\mathring{\mathcal{D}}_{a} = \swne{e}{a}{\mu} \mathring{\nabla}_{\mu}$ \\
    \hline
    $\square_{\text{dR}}$ & $\mathrm{d}^{\dagger} \mathrm{d} + \mathrm{d} \mathrm{d}^{\dagger}$ & 0 & \\
    $\square_{\text{B}}$ & $-\mathring{\nabla}^{\mu} \mathring{\nabla}_{\mu}$ & 0 & \\
    $\blacksquare_{\text{dR}}$ & $\mathrm{D} \mathrm{D}^{\ddag} + \mathrm{D}^{\ddag} \mathrm{D}$ & 0 & \\
    $\blacksquare_{\text{B}}$ & $-\mathcal{D}^{a} \mathcal{D}_{a}$ & 0 & \\
  \end{tabular}
  \end{ruledtabular}
\end{table}

Let $\Phi$ be a scalar (i.e., without Lorentz indices) $p$-form,
\begin{equation}
  \Phi = \frac{1}{p!} \Phi_{\mu_{1} \cdots \mu_{p}} 
  \mathrm{d} x^{\mu_{1}} \wedge \cdots \wedge \mathrm{d} x^{\mu_{p}}.
\end{equation}
There are at least two wave operators that can conceivably act on $\Phi$.
The \emph{Laplace--de~Rham operator},
\begin{equation}
  \square_{\text{dR}} = \mathrm{d}^{\dag} \mathrm{d} + \mathrm{dd}^{\dag},
\end{equation}
is defined as the anticommutator of the exterior derivative
and the \emph{exterior coderivative}, $\mathrm{d}^{\dagger} = *\mathrm{d}*$
[for dimensions other than four or signatures other than $\left(-+++\right)$,
the definition of $\mathrm{d}^{\dagger}$ must be modified
with a judiciously chosen sign].
This operator satisfies the \emph{Weitzenböck identity}
\begin{equation}
  \square_{\mathrm{dR}} \Phi = \square_{\mathrm{B}} \Phi +
  \Sigma_{a} \left(
    \nwse{\mathring{R}}{a}{b} \wedge \Sigma^{b} \Phi
  \right),
  \label{Eq_Weitzenbock_original}
\end{equation}
where $\square_{\mathrm{B}} = -\mathring{\nabla}^{\mu} \mathring{\nabla}_{\mu}$
is the usual Laplace--Beltrami operator built from the torsion-free
covariant derivative $\mathring{\nabla}_{\mu}$.
While unconventional, writing the Weitzenböck identity as in
eq.~(\ref{Eq_Weitzenbock_original}) proves to be useful
for our purposes and is equivalent to more common approaches.
In words, the Weitzenböck identity states that the difference between
the two wave operators acting on $\Phi$ is related to the curvature
of the manifold and does not involve derivatives of $\Phi$.

By definition, the Laplace--Beltrami operator carries no information about torsion.
Since it is the Riemann curvature two-form, $\mathring{R}^{ab}$, that appears
in the second term on the right-hand side of eq.~(\ref{Eq_Weitzenbock_original}),
this means that eq.~(\ref{Eq_Weitzenbock_original}) has no information at all
about torsion, even when torsion is present in spacetime.
This is consistent with the fact that the Laplace--de~Rham operator
is defined without any reference to torsion.

An example of the usefulness of this construction is provided
by classical electromagnetism on a curved spacetime.
Let $A$ be the electromagnetic potential one-form and $F=\mathrm{d}A$ its
associated field strength two-form.
Maxwell equations in vacuum can be written as
\begin{equation}
  \mathrm{d}^{\dag} F = \mathrm{d}^{\dag} \mathrm{d} A = 0.
\end{equation}
Choosing the Lorenz gauge, $\mathrm{d}^{\dag}A=0$, we can use eq.~(\ref{Eq_Weitzenbock_original}) to find
\begin{equation}
  \square_{\mathrm{B}} A +
  \Sigma_{a} \nwse{\mathring{R}}{a}{b} \Sigma^{b} A = 0,
  \label{Eq_Wave_Maxwell_Formas}
\end{equation}
or, in standard tensor language,
\begin{equation}
  -\mathring{\nabla}^{\lambda} \mathring{\nabla}_{\lambda} A_{\mu} +
  \mathring{R}_{\mu\nu} A^{\nu} = 0,
\end{equation}
where $\mathring{R}_{\mu\nu}$ is the standard torsionless Ricci tensor.
It is interesting to notice that this result holds even when the background geometry has nonvanishing torsion.
The electromagnetic field only interacts with the torsionless sector of the geometry.
The same happens with all YM gauge bosons:
they only can interact with the torsionless sector of the geometry.%
\footnote{See footnote~\ref{ft:YM} on page~\pageref{ft:YM}.}

Extending the de~Rham definition of the wave operator for the case
of a $p$-form with $m$ free Lorentz indices, such as
\begin{equation}
  \Psi^{a_{1} \cdots a_{m}} = \frac{1}{p!}
  \Psi^{a_{1} \cdots a_{m}}{}_{\mu_{1} \cdots \mu_{p}}
  \mathrm{d} x^{\mu_{1}} \wedge \cdots \wedge \mathrm{d} x^{\mu_{p}},
\end{equation}
is nontrivial when the geometry has nonvanishing torsion.
As a first step, one may be inclined to define the de~Rham Lorentz-covariant coderivative as $\mathrm{D}^{\dag} = \ast \mathrm{D} \ast$,
in perfect analogy with $\mathrm{d}^{\dagger} = \ast \mathrm{d} \ast$.
We find, however, that a more useful definition is
\begin{equation}
  \mathrm{D}^{\ddag} = - \Sigma^{a} \mathrm{D} \Sigma_{a}.
  \label{Eq_Def_Covariant_Coderivative}
\end{equation}
This is equivalent to the first definition when torsion is zero,
\begin{equation}
 \ast \mathrm{\mathring{D}} \ast =
 -\Sigma^{a} \mathrm{\mathring{D}} \Sigma_{a}, 
\end{equation}
but not in general.
What makes definition~(\ref{Eq_Def_Covariant_Coderivative}) useful
is that the wave operator built from it satisfies
a generalized version of the Weitzenböck identity~(\ref{Eq_Weitzenbock_original}).

Let us define the \emph{generalized Laplace--de~Rham operator} as
\begin{equation}
  \blacksquare_{\text{dR}} =
  \mathrm{DD}^{\ddag} + \mathrm{D}^{\ddag} \mathrm{D}.
  \label{eq:blackdR}
\end{equation}
It is possible to prove that $\blacksquare_{\text{dR}}$
satisfies the following generalized Weitzenböck identity:
\begin{align}
  \blacksquare_{\text{dR}} \Phi^{a_{1} \cdots a_{m}} & =
  \blacksquare_{\text{B}} \Phi^{a_{1} \cdots a_{m}} +
  \Sigma_{c} \mathrm{D}^{2} \Sigma^{c} \Phi^{a_{1} \cdots a_{m}}
  \nonumber \\ & =
  \blacksquare_{\text{B}} \Phi^{a_{1} \cdots a_{m}} + \Sigma_{c}
  \left( R^{c}{}_{b} \Sigma^{b} \Phi^{a_{1} \cdots a_{m}} + \right.
  \nonumber \\ & +
  R^{a_{1}}{}_{b} \Sigma^{c} \Phi^{ba_{2} \cdots a_{m}} + \cdots +
  \nonumber \\ & + \left.
    R^{a_{m}}{}_{b} \Sigma^{c} \Phi^{a_{1} \cdots a_{m-1}b}
  \right).
  \label{Eq_Weitzenboeck_Generalizado} 
\end{align}
In eq.~(\ref{Eq_Weitzenboeck_Generalizado}) we have introduced
the generalized Laplace--Beltrami operator
\begin{equation}
  \blacksquare_{\text{B}} = -\mathcal{D}^{a} \mathcal{D}_{a},
\end{equation}
where
\begin{equation}
  \mathcal{D}_{a} = \Sigma_{a} \mathrm{D} + \mathrm{D} \Sigma_{a}. 
\end{equation}

In the torsionless case, the operator
$\mathcal{\mathring{D}}_{a} =
 \Sigma_{a} \mathrm{\mathring{D}} + \mathrm{\mathring{D}} \Sigma_{a}$
can be shown to satisfy
$\mathcal{\mathring{D}}_{a} = e_{a}{}^{\mu} \mathring{\nabla}_{\mu}$,
meaning that it matches the usual torsionless covariant derivative $\mathring{\nabla} = \partial + \mathring{\Gamma}$,
and the standard Weitzenböck identity~(\ref{Eq_Weitzenbock_original})
is recovered.

Some useful properties satisfied by $\mathcal{D}_{a}$ are
\begin{align}
 \mathcal{D}_{a} \left( \alpha \wedge \beta \right) & =
 \mathcal{D}_{a} \alpha \wedge \beta +
 \alpha \wedge \mathcal{D}_{a} \beta,
 \label{eq:LeibnizCalD}
 \\
 \left[ \Sigma_{a}, \mathcal{D}_{b} \right] & = -
 \left( \Sigma_{ab} T^{c} \right) \Sigma_{c},
 \label{Eq_Comm_Sigma_D}
 \\
 \left[ \mathcal{D}_{a}, \mathcal{D}_{b} \right] & =
 \mathrm{D}^{2} \Sigma_{ab} + \Sigma_{ab} \mathrm{D}^{2} + \Sigma_{a} \mathrm{D}^{2} \Sigma_{b} - \Sigma_{b} \mathrm{D}^{2} \Sigma_{a} +
 \nonumber \\ & -
 \left( \mathrm{D} \Sigma_{ab} T^{c} \right) \wedge \Sigma_{c} -
 \left( \Sigma_{ab} T^{c} \right) \mathcal{D}_{c},
 \label{Eq_Comm_DD}
\end{align}
where $\alpha$ is a $p$-form and $\beta$ is a $q$-form.
In particular, eq.~(\ref{eq:LeibnizCalD}) implies that $\mathcal{D}_{a}$
obeys Leibniz's rule without any correcting signs.

From the above discussion, it seems clear that in order to have waves interacting with torsion, it is necessary for the field to have free Lorentz indices. This is precisely the case of gravitational waves in the Horndeski case, as we shall see in the next section.


\section{Gravitational waves and torsional modes}
\label{sec:GWTM}

\subsection{Linear perturbations for a theory of gravity in the first-order formalism}

Let us consider a background geometry described by $\bar{e}^{a}$, $\bar{\omega}^{ab}$ and $\bar{\phi}$. Linear perturbations\footnote{It is very important to remember that in order to study cases of astrophysical interest, it is necessary to go at least to second-order in the perturbations of curvature. In the current article we are not interested in modelling a particular phenomena, but just studying how gravitational waves could interact with torsion at first-order. Detailed calculations to second-order for particular astrophysical situations will be presented elsewhere.} around this background are described by
\begin{align}
 \bar{e}^{a} \to e^{a} & =
 \bar{e}^{a} + \frac{1}{2} h^{a},
 \label{Eq_pert_e}
 \\
 \bar{\omega}^{ab} \to \omega^{ab} & =
 \bar{\omega}^{ab} + u^{ab},
 \label{Eq_pert_omega}
 \\
 \bar{\phi} \to \phi & =
 \bar{\phi} + \varphi,
 \label{Eq_pert_phi}
\end{align}
where we have introduced the one-forms
$h^{a} = \nwse{h}{a}{b} \bar{e}^{b}$ and
$u^{ab} = \nwse{u}{ab}{c} \bar{e}^{c}$,
and the zero-form $\varphi$.

The linear perturbation of the metric reads
\begin{align}
  g & = \eta_{ab} e^{a} \otimes e^{b}
  \nonumber \\ & =
  \eta_{ab} \left(
    \bar{e}^{a} + \frac{1}{2} \nwse{h}{a}{c} \bar{e}^{c}
  \right)
  \otimes \left(
    \bar{e}^{b} + \frac{1}{2} \nwse{h}{b}{d} \bar{e}^{d}
  \right),
  \nonumber \\ & =
  \eta_{ab} \bar{e}^{a} \otimes \bar{e}^{b} +
  \frac{1}{2} \left(
    h_{ab} + h_{ba}
  \right)
  \bar{e}^{a} \otimes \bar{e}^{b},
  \nonumber \\ & =
  \left(
    \bar{g}_{\mu\nu} + h_{\mu\nu}^{+}
  \right)
  \mathrm{d} x^{\mu} \otimes \mathrm{d} x^{\nu},
\end{align}
where
\begin{equation}
  h_{ab}^{\pm} = \frac{1}{2} \left( h_{ab} \pm h_{ba} \right)
\end{equation}
are the symmetric and antisymmetric parts of $h_{ab}$.
The standard theory of gravitational waves is formulated just in terms of $h_{\mu \nu}^{+}$,
because it is possible to show that the antisymmetric part, $h_{ab}^{-}$,
amounts to nothing more than an infinitesimal local Lorentz transformation.
Since the Horndeski Lagrangian~(\ref{Ec_Horndeski_Formas}) is locally Lorentz invariant,
it is possible to gauge away that piece and to keep only the symmetric part.
Therefore, from now on we will just assume that $h_{ab}$ is symmetric, i.e., $h_{ba} = h_{ab}$.

In standard general relativity, the perturbation in the geometry is described in terms of $h_{\mu\nu}$ alone,
since the perturbation in the connection depends on the $h_{\mu \nu}$ through the torsionless condition.
When considering nonvanishing torsion, the vierbein and the spin connection correspond to independent degrees of freedom.
Therefore, the perturbation one-forms $h^{a}$ and $u^{ab}$ must be independent, too.
Here we show that it is always possible to split the linear perturbation one-form $u^{ab}$ in two pieces,
one carrying all the dependency on $h^{a}$ and one completely independent from it
(and associated, of course, to linear perturbations in the torsion).


Let us begin by writing down the two-form torsion as $T^{a} = \mathrm{D} e^{a}$.
Its linear perturbation under eqs.~(\ref{Eq_pert_e})--(\ref{Eq_pert_phi}) is given by
\begin{align}
  \bar{T}^{a} \to T^{a} & =
  \bar{T}^{a} + \frac{1}{2} \mathrm{\bar{D}} h^{a} +
  \nwse{u}{a}{b} \wedge \bar{e}^{b},
  \nonumber \\ & =
  \bar{T}^{a} + \frac{1}{2} \bring{\mathrm{D}} h^{a} +
  \frac{1}{2} \nwse{\bar{\kappa}}{a}{b} \wedge h^{b} +
  \nwse{u}{a}{b} \wedge \bar{e}^{b},
  \label{Eq_Pert_T_De}
\end{align}
where $\bring{\mathrm{D}}$ denotes the exterior covariant derivative
with respect to the torsionless piece of the background spin connection, $\bring{\omega}^{ab}$.

On the other hand, torsion may be also written in terms of the contorsion one-form, $\kappa^{ab}$,
as $T^{a} = \nwse{\kappa}{a}{b} \wedge e^{b}$. Its linear perturbation reads
\begin{equation}
  \bar{T}^{a} \to T^{a} =
  \bar{T}^{a} + \frac{1}{2} \nwse{\kappa}{a}{b} \wedge h^{b} +
  \nwse{q}{a}{b} \wedge \bar{e}^{b},
  \label{Eq_Pert_T_ke}
\end{equation}
where $q^{ab}$ stands for the linear perturbation in the contorsion, i.e.,
$\bar{\kappa}^{ab} \to \kappa^{ab} = \bar{\kappa}^{ab} + q^{ab}$.

Equations~(\ref{Eq_Pert_T_De}) and~(\ref{Eq_Pert_T_ke}) may seem contradictory at first sight,
since one of them includes derivatives of $h^{a}$ and the other doesn't.
There is no contradiction, though; to see this, one need only notice that $u^{ab}$ must be of the form
\begin{equation}
  u^{ab} = \mathring{u}^{ab} + q^{ab},
  \label{eq:uuq}
\end{equation}
where
\begin{equation}
 \frac{1}{2} \bring{\mathrm{D}} h^{a} +
 \nwse{\mathring{u}}{a}{b} \wedge \bar{e}^{b}=0.
 \label{Eq_u_torsionless}
\end{equation}
Using eqs.~(\ref{eq:uuq}) and~(\ref{Eq_u_torsionless}) in eq.~(\ref{Eq_Pert_T_De}),
the apparent contradiction is resolved.

In order to avoid an algebraic nightmare where both $\bring{\mathrm{D}}$
and $\bar{\mathrm{D}}$ derivatives get mixed together, it is convenient to
define the new perturbation variables
\begin{align}
  \mathcal{U}_{ab} & = \mathring{u}_{ab} -
  \frac{1}{2} \left[
    \bar{\Sigma}_{a} \left(
      \bar{\kappa}_{bc} \wedge h^{c}
    \right)
    - \bar{\Sigma}_{b} \left(
      \bar{\kappa}_{ac} \wedge h^{c}
    \right)
  \right],
  \label{eq:Udef}
  \\
  \mathcal{V}_{ab} & = q_{ab} +
  \frac{1}{2} \left[
    \bar{\Sigma}_{a} \left(
      \bar{\kappa}_{bc} \wedge h^{c}
    \right)
    - \bar{\Sigma}_{b} \left(
      \bar{\kappa}_{ac} \wedge h^{c}
    \right)
  \right].
  \label{eq:Vdef}
\end{align}
These clearly satisfy
\begin{equation}
  u^{ab} = \mathring{u}^{ab} + q^{ab}
  = \mathcal{U}^{ab} + \mathcal{V}^{ab}.
\end{equation}
Using the fact that torsion and contorsion are related by
\begin{equation}
  \kappa_{ab} = \frac{1}{2} \left(
    \Sigma_{a} T_{b} - \Sigma_{b} T_{a} + e^{c} \Sigma_{ab} T_{c}
  \right),
\end{equation}
one can show that eq.~(\ref{Eq_u_torsionless}) becomes
\begin{equation}
  \frac{1}{2} \bar{\mathrm{D}} h^{a} + \nwse{\mathcal{U}}{a}{b} \wedge \bar{e}^{b} +
  \frac{1}{2} \bar{\Sigma}^{a} \left( h_{b} \wedge \bar{T}^{b} \right) = 0,
  \label{Eq_U_torsion_condition}
\end{equation}
and that eq.~(\ref{Eq_Pert_T_ke}) becomes the torsion linear perturbation equation
\begin{equation}
  \bar{T}^{a} \to T^{a} =
  \bar{T}^{a} + \nwse{\mathcal{V}}{a}{b} \wedge \bar{e}^{b} -
  \frac{1}{2} \bar{\Sigma}^{a} \left( h_{b} \wedge \bar{T}^{b} \right).
 \label{Eq_pert_Torsion}
\end{equation}

From eq.~(\ref{Eq_U_torsion_condition}), and after some algebra,
it is possible to get a closed expression for $\mathcal{U}^{ab}$,
\begin{equation}
  \mathcal{U}^{ab} = -\frac{1}{2} \left(
    \bar{\Sigma}^{a} \bar{\mathrm{D}} h^{b} - \bar{\Sigma}^{b} \bar{\mathrm{D}} h^{a}
  \right).
  \label{Eq_U=Dh}
\end{equation}

The linear perturbation of the Lorentz curvature reads simply
\begin{equation}
  \bar{R}^{ab} \to R^{ab} =
  \bar{R}^{ab} + \bar{\mathrm{D}} \left(
    \mathcal{U}^{ab} + \mathcal{V}^{ab}
  \right).
  \label{Eq_pert_Lorentz_curvature}
\end{equation}

We have thus been able to split the Lorentz connection perturbation, $u^{ab}$,
in two parts, $\mathcal{U}^{ab}$ and $\mathcal{V}^{ab}$, such that
$\mathcal{U}^{ab}$ is completely determined by the vierbein perturbation
[via eq.~(\ref{Eq_U=Dh})], and the linear perturbation on the torsion
depends only on $\mathcal{V}^{ab}$ [cf.~eq.~(\ref{Eq_pert_Torsion})].
The Lorentz curvature perturbation, on the other hand, depends on both parts
of the Lorentz connection perturbation, as shown in eq.~(\ref{Eq_pert_Lorentz_curvature}).

Finally, one can show that the scalar field ``curvature''
$Z^{a} = \Sigma^{a} \mathrm{d} \phi$ becomes
\begin{equation}
  \bar{Z}^{a} \to Z^{a} =
  \bar{Z}^{a} + \bar{\Sigma}^{a} \mathrm{d} \varphi -
  \frac{1}{2} \nwse{h}{a}{b} \bar{Z}^{b}.
  \label{Eq_pert_Z}
\end{equation}

\subsection{Gravitational Waves and Torsion in Horndeski's Theory}
\label{sec:GWEH}

As we have seen in section~\ref{sec:1OFHT}, nonminimal couplings
and second-order derivatives terms in the Horndeski lagrangian
are sources of torsion.
In this general case, torsion propagates through the ``contorsional mode''
$\mathcal{V}^{ab}$ and the background torsion $\bar{T}^{a}$ interacts
with the metric modes $h^{a}$.
However, our intuition may lead us to believe that the Einstein--Hilbert (EH) term
can give rise only to the wave equation and interactions of $h^{a}$
with the background curvature, as in the standard torsionless case.
That is not the case.
As we shall see, even the EH term gives rise to both,
metrical modes interacting with the background torsion and propagating torsional modes.

In order to see this, let us consider a Lagrangian in the Horndeski family of the form
\begin{equation}
  \mathcal{L}^{\left( 4 \right)} \left( e, \omega, \phi \right) =
  \mathcal{L}_{\text{EH}}^{\left( 4 \right)} +
  \left( \text{other terms} \right),
\end{equation}
where these ``other terms'' are the ones giving rise to torsion
through nonminimal couplings and/or second-order derivatives of $\phi$.
The EH four-form term is given by
\begin{equation}
  \mathcal{L}_{\text{EH}}^{\left( 4 \right)} \left( e, \omega \right) =\
  \frac{1}{4\kappa_{4}} \epsilon_{abcd} R^{ab} \wedge e^{c} \wedge e^{d},
\end{equation}
and therefore the field equations take the form
\begin{align}
  \delta_{e} \mathcal{L}^{\left( 4 \right)} \left( e, \omega, \phi \right) & =
  \delta_{e} \mathcal{L}_{\text{EH}}^{\left( 4 \right)} +
  \delta_{e} \left( \text{other terms} \right) = 0,
  \label{Eq_motion_EH+others_e} \\
  \delta_{\omega} \mathcal{L}^{\left( 4 \right)} \left( e, \omega, \phi \right) & =
  \delta_{\omega} \mathcal{L}_{\text{EH}}^{\left( 4 \right)} +
  \delta_{\omega} \left( \text{other terms} \right) = 0,
  \label{Eq_motion_EH+others_w} \\
  \delta_{\phi} \mathcal{L}^{\left( 4 \right)} \left( e, \omega, \phi \right) & =
  \delta_{\phi} \left( \text{other terms} \right) = 0,
  \label{Eq_motion_EH+others_phi}
\end{align}
where
\begin{align}
  \delta_{e} \mathcal{L}_{\text{EH}}^{\left( 4 \right)} \left( e, \omega \right) & =
  \frac{1}{2\kappa_{4}} \epsilon_{abcd} R^{ab} \wedge e^{c} \wedge \delta e^{d},
  \\
  \delta_{\omega} \mathcal{L}_{\text{EH}}^{\left( 4 \right)} \left( e, \omega \right) & =
  \frac{1}{2\kappa_{4}} \epsilon_{abcd} \delta \omega^{ab} \wedge T^{c} \wedge e^{d}.
\end{align}

We consider now a background configuration
$\bar{e}^{a}$, $\bar{\omega}^{ab}$, $\bar{\phi}$ satisfying the
field equations~(\ref{Eq_motion_EH+others_e})--(\ref{Eq_motion_EH+others_phi})
and linear perturbations around it as in eqs.~(\ref{Eq_pert_e})--(\ref{Eq_pert_phi}).
When doing this, the result reads
\begin{align}
  \mathcal{G} +
  \frac{1}{4\kappa_{4}} \epsilon_{abcd} \bar{R}^{ab} \wedge h^{c} \wedge \delta e^{d}
  & + \nonumber \\ +
  \frac{1}{2\kappa_{4}} \epsilon_{abcd} \bar{\mathrm{D}} \mathcal{V}^{ab} \wedge
  \bar{e}^{c} \wedge \delta e^{d}
  & + \nonumber \\ + 
  \left( \text{linear perturbations of other terms} \right)
  & = 0,
  \label{Eq_perturbations_EH+other}
\end{align}
where the four-form $\mathcal{G}$ is given by
\begin{equation}
  \mathcal{G} = \frac{1}{2\kappa_{4}} \epsilon_{abcd}
  \bar{\mathrm{D}} \mathcal{U}^{ab} \wedge \bar{e}^{c} \wedge \delta e^{d}.
  \label{eq:Gdef}
\end{equation}
The $\mathcal{G}$-term generates a gravitational wave
described by the generalized wave operator
[cf.~eq.(\ref{eq:blackdR})]
$\blacksquare_{\text{dR}} = \mathrm{D}^{\ddag} \mathrm{D} + \mathrm{DD}^{\ddag}$
coupled with torsion in a nontrivial way,
in strong contrast with the example of eq.~(\ref{Eq_Wave_Maxwell_Formas}).

Using eq.~(\ref{Eq_U=Dh}) in eq.~(\ref{eq:Gdef}), we get
\begin{align}
  \mathcal{G} & = -\frac{1}{4\kappa_{4}} \bar{\ast} \left\{
    \bar{\mathcal{D}}^{a} \bar{\mathcal{D}}_{a} h_{d} -
    \bar{\mathcal{D}}_{a} \bar{\mathcal{D}}_{d} h^{a} +
  \right.
  \nonumber \\ &
  -\bar{e}^{c} \bar{\mathcal{D}}_{c} \left(
    \bar{\Sigma}_{d} \bar{\mathcal{D}}_{a} h^{a} - \bar{\mathcal{D}}_{d} h
  \right)
  + \nonumber \\ &
  -\frac{1}{2} \left[
    \bar{\Sigma}_{b} \left(
      \bar{\mathcal{D}}^{a} \bar{\mathcal{D}}_{a} h^{b} -
      \bar{\mathcal{D}}_{a} \bar{\mathcal{D}}^{b} h^{a}
    \right)
    +
  \right.
  \nonumber \\ &
  \left.
    \left.
      -\bar{\mathcal{D}}_{b} \left(
        \bar{\Sigma}^{b} \bar{\mathcal{D}}_{a} h^{a} - \bar{\mathcal{D}}^{b} h
      \right)
    \right]
    \bar{e}_{d}
  \right\}
  \wedge \delta e^{d},
  \label{Eq_G(h)}
\end{align}
where $\bar{\ast}$ stands for the Hodge dual under the background metric structure
associated to $\bar{e}^{a}$,
$\bar{\mathcal{D}}^{a} = \bar{\Sigma}^{a} \bar{\mathrm{D}} +
 \bar{\mathrm{D}} \bar{\Sigma}^{a}$,
with the operator
$\bar{\Sigma}^{a} = -\bar{\ast} \left( \bar{e}^{a} \wedge \bar{\ast} \right.$,
and $h = \bar{\Sigma}_{a} h^{a}$
(see Table~\ref{tab:dbag} for a summary of the different derivatives defined in this paper).

Let $\zeta$ be a vector field.
An infinitesimal Lie dragging $1 - \pounds_{\zeta}$ generated by $\zeta$
on the background geometry corresponds to
\begin{equation}
  h_{ab} \to h_{ab}^{\prime} =
  h_{ab} - \left(
    \bar{\mathcal{D}}_{a} \zeta_{b} + \bar{\mathcal{D}}_{b} \zeta_{a}
  \right)
  + \zeta^{c} \bar{\Sigma}_{c} \left(
    \bar{\Sigma}_{a} \bar{T}_{b} + \bar{\Sigma}_{b} \bar{T}_{a}
  \right).
  \label{Eq_h_Lie_drag_symm}
\end{equation}
Performing the standard change of variable
$h^{a} \to \tilde{h}^{a}$,
\begin{equation}
  h^{a} = \tilde{h}^{a} - \frac{1}{2} \bar{e}^{a} \tilde{h},
\end{equation}
where $\tilde{h}$ is the trace of the new Lorentz vector one-form variable $\tilde{h}^{a}$,
it is possible to prove that under~(\ref{Eq_h_Lie_drag_symm}), the ``divergence''
$\bar{\mathcal{D}}_{a} \tilde{h}^{a} = \left(
   \bar{\Sigma}_{a} \bar{\mathrm{D}} + \bar{\mathrm{D}} \bar{\Sigma}_{a}
 \right)
 \tilde{h}^{a}$
transforms as
$\bar{\mathcal{D}}_{a} \tilde{h}^{a} \to
 \bar{\mathcal{D}}_{a} \tilde{h}'^{a}$,
where
$\bar{\mathcal{D}}_{a} \tilde{h}'^{a}$
is given by
\begin{align}
  \bar{\mathcal{D}}_{a} \tilde{h}'^{a} & = \left[
    -\bring{\mathcal{D}}_{a} \bring{\mathcal{D}}^{a} \zeta_{b} +
    \bar{\mathcal{D}}_{a} \bar{\Sigma}_{b} \tilde{h}^{a} -
    \bar{\Sigma}_{ab} \nwse{\bring{R}}{a}{c} \zeta^{c} +
  \right.
  \nonumber \\ &
  -\bar{\Sigma}^{ac} \bar{T}_{a} \left(
    \bring{\mathcal{D}}_{c} \zeta_{b} +
    \bring{\mathcal{D}}_{b} \zeta_{c} -
    \eta_{cb} \bring{\mathcal{D}}_{p} \zeta^{p}
  \right)
  + \nonumber \\ & \left.
    + \bar{\Sigma}_{cb} \bar{T}_{a} \bar{\Sigma}^{c} \tilde{h}^{a}
  \right]
  \bar{e}^{b},
  \label{Eq_Dh=gauge}
\end{align}
with
$\bring{\mathcal{D}}_{a} = \bar{\Sigma}_{a} \bring{\mathrm{D}} +
\bring{\mathrm{D}} \bar{\Sigma}_{a}$.
This means we can always choose the ``Lorenz gauge'' 
\begin{equation}
  \bar{\mathcal{D}}_{a} \tilde{h}^{a} = 0
\end{equation}
with a vector field $\zeta$ such that the right-hand side
of eq.~(\ref{Eq_Dh=gauge}) vanishes.

Choosing this gauge and using eqs.~(\ref{Eq_Comm_Sigma_D})--(\ref{Eq_Comm_DD}),
it is possible to recast eq.~(\ref{Eq_perturbations_EH+other})
in terms of $\tilde{h}^{a}$ as
\begin{align}
  \bar{\blacksquare}_{\text{dR}} \tilde{h}^{d} + \bar{\Sigma}_{ad} \left(
    \nwse{\bar{R}}{a}{b} \wedge \tilde{h}^{b}
  \right)
  & + \nonumber \\
  - \left\{
    A_{d} + B_{d} + \frac{1}{2} \bar{e}_{d} \left[
      C - \bar{\Sigma}_{c} \left( A^{c} + B^{c} \right)
    \right]
  \right\}
  & + \nonumber \\
  + \epsilon_{abcd} \bar{\ast} \left(
    \bar{R}^{ab} \wedge h^{c} +
    2 \bar{\mathrm{D}} \mathcal{V}^{ab} \wedge \bar{e}^{c}
  \right)
  & + \nonumber \\
  + \left( \text{linear perturbations of other terms} \right) & = 0,
  \label{Eq_G_final}
\end{align}
where $\bar{\blacksquare}_{\text{dR}}\tilde{h}_{d}$
is given by the generalized Weitzenböck identity~(\ref{Eq_Weitzenboeck_Generalizado}),
\begin{equation}
  \bar{\blacksquare}_{\text{dR}} \tilde{h}_{a} =
  -\bar{\mathcal{D}}^{b} \bar{\mathcal{D}}_{b} \tilde{h}_{a} +
  \bar{\Sigma}_{b} \left(
    \nwse{\bar{R}}{b}{c} \bar{\Sigma}^{c} \tilde{h}_{a} -
    \nwse{\bar{R}}{c}{a} \bar{\Sigma}^{b} \tilde{h}_{c}
  \right),
\end{equation}
and $A_{a},$ $B_{a}$ and $C$ are the torsional terms
\begin{align}
  A_{a} & = \left( \bar{\Sigma}_{ca} \bar{T}_{b} \right)
  \mathcal{\bar{D}}^{b} \tilde{h}^{c} +
  \tilde{h}^{bc} \mathrm{\bar{D}} \bar{\Sigma}_{ca} \bar{T}_{b},
  \label{Eq_Tor_Am}
  \\
  B_{a} & = \left( \bar{\Sigma}_{c} \bar{T}_{b} \right)
  \bar{\Sigma}^{b} \left[
    \mathcal{\bar{D}}_{a} \left( \tilde{h} \bar{e}^{c} \right) -
    \mathcal{\bar{D}}^{c} \left( \tilde{h} \bar{e}_{a} \right)
  \right]
  + \nonumber \\ & +
  \frac{1}{2} \left\{
    \tilde{h} \mathrm{\bar{D}}^{\ddag} \bar{T}_{a} +
    \bar{\Sigma}^{b} \left[
      \mathrm{\bar{D}} \left( \tilde{h} \bar{\Sigma}_{a} \bar{T}_{b} \right) -
      \bar{T}_{b} \bar{\Sigma}_{a} \mathrm{\bar{D}} \tilde{h}
    \right]
  \right\},
  \\
  C & = \mathcal{\bar{D}}^{c} \left( \tilde{h}^{ab} \bar{\Sigma}_{bc} \bar{T}_{a} \right) +
  \left( \bar{\Sigma}_{bc} \bar{T}_{a} \right)
  \bar{\Sigma}^{a} \mathcal{\bar{D}}^{c} \tilde{h}^{b}.
  \label{Eq_Tor_C}
\end{align}

The equation for the propagation of linear perturbations is found by replacing
eq.~(\ref{Eq_G_final}) in eq.~(\ref{Eq_perturbations_EH+other}).
Doing so, we observe that in the context of nonvanishing torsion:
\begin{itemize}
  \item The metric wave $\tilde{h}_{ab}$ couples to both,
        the background torsion \emph{and} the background curvature.
  \item The metric wave $\tilde{h}_{ab}$ couples to an independent
        propagating torsion wave mode, $\mathcal{V}^{ab}$.
  \item Some of the coupling between $\tilde{h}_{ab}$ and the background torsion
        occurs through the trace $\tilde{h}$. All this dependence has been ``packed''
        in the Lorentz-vector one-form $B_{a}$, but the important point is that the
        ``traceless'' variable $\tilde{h}_{ab}$ no longer leads to equations
        without the trace $\tilde{h}$.
\end{itemize}

\subsection{Gravitational Waves and generic terms of the Horndeski Lagrangian}
In section~\ref{sec:GWEH} we showed that the EH term in the Horndeski Lagrangian
can produce gravitational waves interacting with the background torsion and
propagating torsional modes.
In this section we highlight those other terms in the Horndeski family that can
lead to similar behavior.

Generic terms will couple $\tilde{h}^{a}$, $\mathcal{V}^{ab}$ and $\varphi$
with the background curvature $\bar{R}^{ab}$ and torsion $\bar{T}^{a}$,
but only some very specific terms will contribute with second-order wave-like
operators on the metric mode ($\partial^{2}\tilde{h}^{a}$ terms),
and first-order operators acting on the torsional mode ($\partial \mathcal{V}^{ab}$ terms).

For the sake of simplicity, let us focus on linear perturbations of the vierbein
and spin connection, leaving the scalar field unchanged (i.e., $\varphi = 0$).
In this case, the linear perturbations on
$e^{a}$, $R^{ab}$, $T^{a}$, $Z^{a}$, $\theta^{a}$ and $\pi^{a}$
(the fundamental ingredients of the field equations) read
\begin{align}
  \bar{e}^{a} \to e^{a} & =
  \bar{e}^{a} + \frac{1}{2} h^{a},
  \\
  \bar{R}^{ab} \to R^{ab} & =
  \bar{R}^{ab} + \bar{\mathrm{D}} \mathcal{V}^{ab} -
  \frac{1}{2} \bar{\mathrm{D}} \left(
    \bar{\Sigma}^{a} \bar{\mathrm{D}} h^{b} -
    \bar{\Sigma}^{b} \bar{\mathrm{D}} h^{a}
  \right),
  \\
  \bar{T}^{a} \to T^{a} & =
  \bar{T}^{a} + \nwse{\mathcal{V}}{a}{b} \wedge \bar{e}^{b} -
  \frac{1}{2} \bar{\Sigma}^{a} \left(
    h_{b} \wedge \bar{T}^{b}
  \right),
  \\
  \bar{Z}^{a} \to Z^{a} & =
  \bar{Z}^{a} - \frac{1}{2} \nwse{h}{a}{b} \bar{Z}^{b},
  \\
  \bar{\theta}^{a} \to \theta^{a} & =
  \bar{\theta}^{a} - \frac{1}{2} \nwse{h}{a}{b} \bar{\theta}^{b},
  \\
  \bar{\pi}^{a} \to \pi^{a} & =
  \bar{\pi}^{a} - \frac{1}{2} \nwse{h}{a}{b} \bar{\pi}^{b} +
  \\ & + \left[
    \mathcal{V}^{ab} - \frac{1}{2} \left(
      \bar{\Sigma}^{a} \bar{\mathrm{D}} h^{b} -
      \bar{\Sigma}^{b} \bar{\mathrm{D}} h^{a} -
      \bar{\mathrm{D}} h^{ab}
    \right)
  \right]
  \bar{Z}_{b}.
\end{align}

In the above equations, only the perturbation of the Lorentz curvature
includes second-order derivatives of $h^{a}$
(through the operator
$\bar{\mathrm{D}} \bar{\Sigma}^{a} \bar{\mathrm{D}}$)
and first-order derivatives of the torsional perturbation
$\bar{\mathrm{D}} \mathcal{V}^{ab}$.
Given that $\mathrm{d}^{2} = 0$, and that the Hodge operator appears in the Lagrangian
only through $\Sigma^{a} = -\ast e^{a} \wedge \left( \ast \right.$,
we find that in the equations of motion $\mathcal{E}_{ab} = 0$ and $\mathcal{E}_{a} = 0$
[cf.~eqs.~(\ref{Eq_Motion_wab}) and~(\ref{Eq_Motion_ea})]
$\partial^{2}\tilde{h}^{a}$ and $\partial\mathcal{V}^{ab}$
can arise only from terms where the curvature is present.

In the Horndeski Lagrangian, this boils down to
(i)~terms where the Lorentz curvature appears explicitly,
\begin{align}
  & \left( F + 2W \right) \epsilon_{abcd} R^{ab} \wedge e^{c} \wedge e^{d},
  \label{Ec_R_e_e} \\
  & \kappa_{3} \epsilon_{abcd} R^{ab} \wedge e^{c} \wedge \theta^{d},
  \label{Ec_R_e_theta} \\
  & \kappa_{1} \epsilon_{abcd} R^{ab} \wedge e^{c} \wedge \pi^{d},
  \label{Ec_R_e_pi}
\end{align}
and (ii)~any terms having two or more $\pi^{a}$'s:
\begin{align}
  & \frac{\partial \kappa_{1}}{\partial X}
    \epsilon_{abcd} \pi^{a} \wedge \pi^{b} \wedge \pi^{c} \wedge e^{d},
    \label{Ec_pi3_e} \\
  & \frac{\partial \kappa_{3}}{\partial X}
    \epsilon_{abcd} \theta^{a} \wedge \pi^{b} \wedge \pi^{c} \wedge e^{d},
    \label{Ec_theta_pi2_e} \\
  & \frac{\partial F}{\partial X}
    \epsilon_{abcd} \pi^{a} \wedge \pi^{b} \wedge e^{c} \wedge e^{d}.
    \label{Ec_pi2_e2}
\end{align}

These three last terms produce curvature in the field equations
through the Bianchi identity,
$\mathrm{D} \pi^{a} = \mathrm{D}^{2} Z^{a} = \nwse{R}{a}{b} Z^{b}$.
This can most easily be seen by considering the dependence of $Z^{a} = \Sigma^{a} \mathrm{d} \phi$
on the vierbein and integrating by parts [see, e.g., eq.~(\ref{Eq_Sa})].

As an example, let us consider the term
\begin{equation}
  \mathcal{L}_{\theta} =
  \frac{1}{2} \epsilon_{abcd} R^{ab} \wedge e^{c} \wedge \theta^{d}.
\end{equation}
The variation of $\mathcal{L}_{\theta}$ under an infinitesimal change in the vierbein reads
\begin{equation}
  \delta_{e} \mathcal{L}_{\theta} = \left[
    \frac{1}{2} \epsilon_{abcd} R^{ab} \wedge \theta^{c} + \Sigma^{a} \left(
      \mathcal{G}_{a} \wedge \theta_{d}
    \right)
  \right]
  \wedge \delta e^{d},
\end{equation}
where the ``Einstein tensor'' three-form $\mathcal{G}_{d}$ is given by
\begin{equation}
  \mathcal{G}_{d} = \frac{1}{2} \epsilon_{abcd} R^{ab} \wedge e^{c}.
\end{equation}

Under linear perturbations, $\delta_{e} \mathcal{L}_{\theta}$ behaves as
\begin{align}
  \delta_{e} \mathcal{L}_{\theta} & =
  \delta_{\bar{e}} \mathcal{\bar{L}}_{\theta} + \left\lbrace
    \bar{\Sigma}^{m} \left(
      \mathcal{W}_{m} \wedge \bar{\theta}_{d} +
      \mathcal{\bar{G}}_{m} \wedge \Upsilon_{d}
    \right)
  \right. + \nonumber \\ & +
  \frac{1}{2} \epsilon_{abcd} \bar{R}^{ab} \wedge \Upsilon^{c}
  + \nonumber \\ & +
  \frac{1}{2} \epsilon_{abcd} \bar{\mathrm{D}} \left[
    \mathcal{V}^{ab} - \frac{1}{2} \left(
      \bar{\Sigma}^{a} \bar{\mathrm{D}} h^{b} -
      \bar{\Sigma}^{b} \bar{\mathrm{D}} h^{a}
    \right)
  \right]
  \wedge \bar{\theta}^{c} +
  \nonumber \\ & \left. -
    \frac{1}{2} h^{mn} \bar{\Sigma}_{n} \left(
      \mathcal{\bar{G}}_{m} \wedge \bar{\theta}_{d}
    \right)
  \right\rbrace
  \wedge \delta e^{d},
\end{align}
where
$\Upsilon^{a} =
 -\frac{1}{2} \nwse{h}{a}{b} \bar{\theta}^{b} +
 \mathrm{d} \bar{\phi} \bar{\Sigma}^{a} \mathrm{d} \varphi +
 \bar{Z}^{a} \mathrm{d} \varphi$
is the linear perturbation of $\theta^{a}$ and $\mathcal{W}_{a}$ is just a
shortcut for the gravitational wave terms we have already seen in eq.~(\ref{Eq_G_final}),
\begin{align}
  \mathcal{W}_{d} & = -\frac{1}{4} \epsilon_{abcd} \bar{\mathrm{D}} \left(
    \bar{\Sigma}^{a} \bar{\mathrm{D}} h^{b} -
    \bar{\Sigma}^{b} \bar{\mathrm{D}} h^{a}
  \right)
  \wedge \bar{e}^{c} +
  \nonumber \\ & +
  \frac{1}{2} \epsilon_{abcd} \left(
    \frac{1}{2} \bar{R}^{ab} \wedge h^{c} +
    \bar{\mathrm{D}} \mathcal{V}^{ab} \wedge \bar{e}^{c}
  \right)
  \nonumber \\ & =
  \frac{1}{4} \bar{\ast} \left[
    \bar{\blacksquare}_{\text{dR}} \tilde{h}_{d} +
    \bar{\Sigma}_{ad} \left(
      \nwse{\bar{R}}{a}{b} \wedge \tilde{h}^{b}
    \right)
  \right]
  + \nonumber \\ & -
  \frac{1}{4} \bar{\ast} \left\lbrace
    A_{d} + B_{d} + \frac{1}{2} \bar{e}_{d} \left[
      C - \bar{\Sigma}_{c} \left(
        A^{c} + B^{c}
      \right)
    \right]
  \right\rbrace
  + \nonumber \\ & +
  \frac{1}{2} \epsilon_{abcd} \left(
    \frac{1}{2} \bar{R}^{ab} \wedge h^{c} +
    \bar{\mathrm{D}} \mathcal{V}^{ab} \wedge \bar{e}^{c}
  \right),
\end{align}
where $A_{a},$ $B_{a}$ and $C$ are the torsion couplings defined in eqs.~(\ref{Eq_Tor_Am})--(\ref{Eq_Tor_C}).

Similar expressions for the linear perturbations of terms like~(\ref{Ec_R_e_theta})--(\ref{Ec_pi2_e2}) can also be found.

Beyond the complicated algebra,
the interesting point is that, quite generally,
every appearance of $\partial^{2}\tilde{h}^{a}$ terms is related with couplings with torsion.
In the context of the Horndeski Lagrangian,
the coupling between $\partial^{2}\tilde{h}^{a}$ terms
and torsion seems to be rather the rule than the exception.

\section{Conclusions}
When YM bosons are described by connections on fiber bundles,
their field strength is given by
$F = \mathrm{d} A + \frac{1}{2} \left[ A, A \right]$,
regardless of the curvature and torsion of the spacetime (basis) manifold.
The YM Lagrangian,
$\mathcal{L}_{\text{YM}} = -\frac{1}{4} \left\langle F \wedge \ast F \right\rangle$,
only has information about the connection $A$
and the background spacetime metric $g_{\mu \nu}$
needed to construct the Hodge $\ast$-operator.
Therefore, YM bosons will be sensitive to the spacetime Riemann (metric) curvature but oblivious to torsion.
Of all Standard Model fields, torsion only interacts, albeit very weakly,
with fermions in the ECSK theory.
Since it is always possible to ``pack'' torsional terms in an effective stress-energy tensor, it may seem tempting to consider torsion as a
dark matter candidate (see, e.g., Ref.~\cite{Belyaev:2016icc}).

Adopting geometry as a solution to the dark matter problem
is an idea with a rich history (see, e.g., Ref.~\cite{Verlinde:2016toy}).
There are, however, at least two potential weak points worth considering:
\begin{itemize}
  \item In the pure ECSK theory, torsion does not propagate in vacuum
        and fermions are its only (very weak) source. Therefore, in order
        to consider the idea seriously it is necessary to look for more general
        theories in $d=4$ and new torsion sources.
  \item The same ``darkness'' of torsion (i.e., its lack of interaction with YM fields)
        that makes the idea attractive also makes it hard to falsify in any foreseeable
        accelerator physics experiment. Therefore it seems appropriate to find a
        torsion-sensitive phenomenon outside of the Standard Model in order to test
        the idea of torsion as dark matter.
\end{itemize}

In this paper we have explored solutions to both of these issues.
Regarding the first point, in section~\ref{sec:1OFHT} 
we take Horndeski's theory and allow it to develop nonzero torsion
by recasting it in Cartan's first-order formalism.
The main result of this exercise is that every nonminimal coupling of the geometry
with $\phi$ and every term in the Lagrangian with second derivatives of $\phi$
are generic sources of torsion.
This was to be expected in the light of previous work, such as
section~1.7.1 of Ref.~\cite{CastellaniSUGRA} on the Brans--Dicke theory
and Ref.~\cite{Tol13} on nonminimal coupling with the Gauss--Bonnett term.
The main novelty of section~\ref{sec:1OFHT} is the development of new mathematical techniques
based on the properties of the $\Sigma^{a}$ operator,
making it accessible to work with the full Horndeski Lagrangian
in first-order formalism and without imposing the torsionless condition.

In section~\ref{sec:GWTM} we explored the idea of using gravitational waves
as a probe for torsion, and in section~\ref{sec:waveops} we introduced the
necessary mathematical tools to address this problem.
In particular, in section~\ref{sec:waveops} we developed
a generalization of the Laplace--de~Rham operator,
$\square_{\text{dR}} = \mathrm{d}^{\dag} \mathrm{d} + \mathrm{dd}^{\dag}$,
to a new operator
$\blacksquare_{\text{dR}} =
\mathrm{D}^{\ddag} \mathrm{D} + \mathrm{DD}^{\ddag}$
which acts covariantly on $p$-forms with Lorentz indices, where 
$\mathrm{D}^{\ddag} = - \Sigma_{a} \mathrm{D} \Sigma^{a}$.
In section~\ref{sec:GWTM} we showed that any Horndeski Lagrangian
that includes the EH term will give rise to gravitational waves,
governed by the $\blacksquare_{\text{dR}}$ operator,
plus new interactions with the background torsion.

The following is an incomplete list of the many problems
that remain open for future work.
\begin{itemize}
  \item It is clear that the Horndeski theorem breaks down
        in the case of nonvanishing torsion:
        there are many new torsional terms which can be added to the Lagrangian
        which give rise only to second-order field equations.
        What is the most general Lagrangian for this case remains as an open problem. 
  \item We have shown in section~\ref{sec:GWTM} that gravitational waves
        interact with the background torsion, and a new torsional mode appears.
        However, the phenomenology of this interaction still remains to be modeled.
        Even further, in any realistic astrophysical scenario it is necessary
        to go up to second-order in perturbations
        (see, e.g., Ref.~\cite{MaggioreGW2008}).
  \item It is not yet clear which, if any, of the Horndeski family members
        generate suitable dark matter profiles. With sufficiently precise
        observations, one may hope to use this information to select
        the most appropriate Lagrangians, or at least rule some of them out.
        The same is true regarding gravitational waves propagation.
        Some ideas have been proposed about this point in Ref.~\cite{Bettoni:2016mij},
        but only for the torsionless case.
  \item The cosmological implications of Horndeski's theory
        have been studied only on particular cases
        (see, e.g., Refs.~\cite{Jackiw:2003pm,Cantcheff:2008qn,Ertem:2009ur,Tol13}).
\end{itemize}


\begin{acknowledgments}
We are grateful to
Antonella Cid,
José~M. Izquierdo,
Patricio Mella,
Julio Oliva,
Patricio Salgado, and
Jorge Zanelli
for many enlightening conversations.
This research was partially funded by Fondecyt grants 
1130653, 1150719 (FI), and
3160437 (OV),
and by Conicyt scholarships
72160340 (FC-T),
21160784 (JB),
21161574 (PM),
and 21161099 (DN)
from the Government of Chile.
ER wishes to thank the German Academic Exchange Service (DAAD)
for financial support, and
Dieter L\"{u}st for his kind hospitality at the
Arnold Sommerfeld Center for Theoretical Physics in Munich.
\end{acknowledgments}

\bibliography{biblio2017.bib}

\end{document}